\documentclass[american,english,aps,prb,preprint,showpacs]{revtex4}
\usepackage[T1]{fontenc}
\usepackage[latin1]{inputenc}
\usepackage{babel}

\makeatletter

\providecommand{\LyX}{L\kern-.1667em\lower.25em\hbox{Y}\kern-.125emX\@}

\makeatother
\begin{document}

\title{Photoluminescence Stokes shift and exciton fine structure in CdTe
nanocrystals }

\author{J.~P\'{e}rez-Conde}

\address{Departamento de F\'{\i}sica, Universidad P\'{u}blica de Navarra,
E-31006 Pamplona, Spain }

\author{A.~K.~Bhattacharjee }

\address{Laboratoire de Physique des Solides, UMR du CNRS, Universit\'{e}
Paris-Sud, 91405 Orsay, France}

\author{M. Chamarro and P. Lavallard}

\address{Groupe de Physique des Solides, Universités Paris VI et VII, CNRS
UMR 7588, 2 Place Jussieu, 75251 Paris, France}

\author{V. D. Petrikov and A. A. Lipovskii}

\address{St. Petersburg State Technical University, Polytechnicheskaja 29,
St. Petersburg 19251, Russian Federation}

\begin{abstract}
The photoluminescence spectra of spherical CdTe nanocrystals with
zincblende structure are studied by size-selective spectroscopic techniques.
We observe a resonant Stokes shift of \( 15 \) meV when the excitation
laser energy is tuned to the red side of the absorption band at \( 2.236 \)
eV. The experimental data are analyzed within a symmetry-based tight-binding
theory of the exciton spectrum, which is first shown to account for
the size dependence of the fundamental gap reported previously in
the literature. The theoretical Stokes shift presented as a function
of the gap shows a good agreement with the experimental data, indicating
that the measured Stokes shift indeed arises from the electron-hole
exchange interaction. 
\end{abstract}
\maketitle
Three-dimensional confinement in semiconductor nanocrystals (NC's),
also called quantum dots (QD's), leads to a discretization of the
electronic energy levels and a blue shift of the absorption edge.
The confinement also enhances the electron-hole direct and exchange
Coulomb interactions because the spatial overlap of the electron and
hole wave functions strongly increases with decreasing size. As a
consequence, the fine structure splittings of the lowest-energy excitonic
states are much larger than in the bulk material. \cite{cn93} Evidence
of such enhanced splittings in the photoluminescence (PL) Stokes shifts
was reported \cite{cg96, er96} in CdSe NC's and more recently \cite{cd98}
in CdS NC's. In both semiconductors, the Stokes shift between the
energy of the resonantly excited first electron-hole state and the
luminescence line is about \( 10 \) meV in NC's of radius \( 15 \)
\AA. It arises from an interplay of the electron-hole exchange interaction
and the crystal-field term in the wurtzite structure CdSe, while the
small spin-orbit interaction plays an important role in CdS. In this
paper we study zincblende CdTe NC's, where the exciton fine structure
is expected to be completely determined by the electron-hole interaction.
We first present an experimental determination of the size-dependent
Stokes shift. A theoretical model based on the tight-binding (TB)
method is then used to analyze the data. It is shown to account for
the available data on the size-dependent absorption gap. Finally,
we present a comparison between the calculated Stokes shift as a function
of the gap and our measured values, showing an excellent agreement.

We report a detailed analysis of the fine structure of the photoluminescence
(PL) spectra observed under resonant excitation. Optical absorption
measurement was performed with a Cary 2300 spectrometer. Non-resonant
photoluminescence excitation was realized with the green line of an
Ar\( ^{+} \) laser at \( 514 \) nm. Resonant PL was obtained with
a Coumarin \( 6 \) dye-laser pumped by the \( 488 \) nm line of
an Ar\( ^{+} \) laser. The PL spectra were analyzed with double substrative
spectrometer which minimizes the rate of diffused light. The spectral
resolution was \( 1 \) meV. 

The sample studied is a phosphate P\( _{2} \)ONa\( _{2} \)O-ZnO-AlF\( _{3} \)-Ga\( _{2} \)O\( _{3} \)
glass containing \( \sim 1.2 \) wt.\% of CdTe. The synthesis of the
material was achieved following a process previously described.\cite{kl95}
The technique of raw batch was applied. Closed crucible technique
was used to prevent decreasing Te concentration through volatility.
After the synthesis and the quenching at \( 300 \)\( ^{o} \)C the
glass samples were properly annealed to grow CdTe nanocrystals of
necessary size.

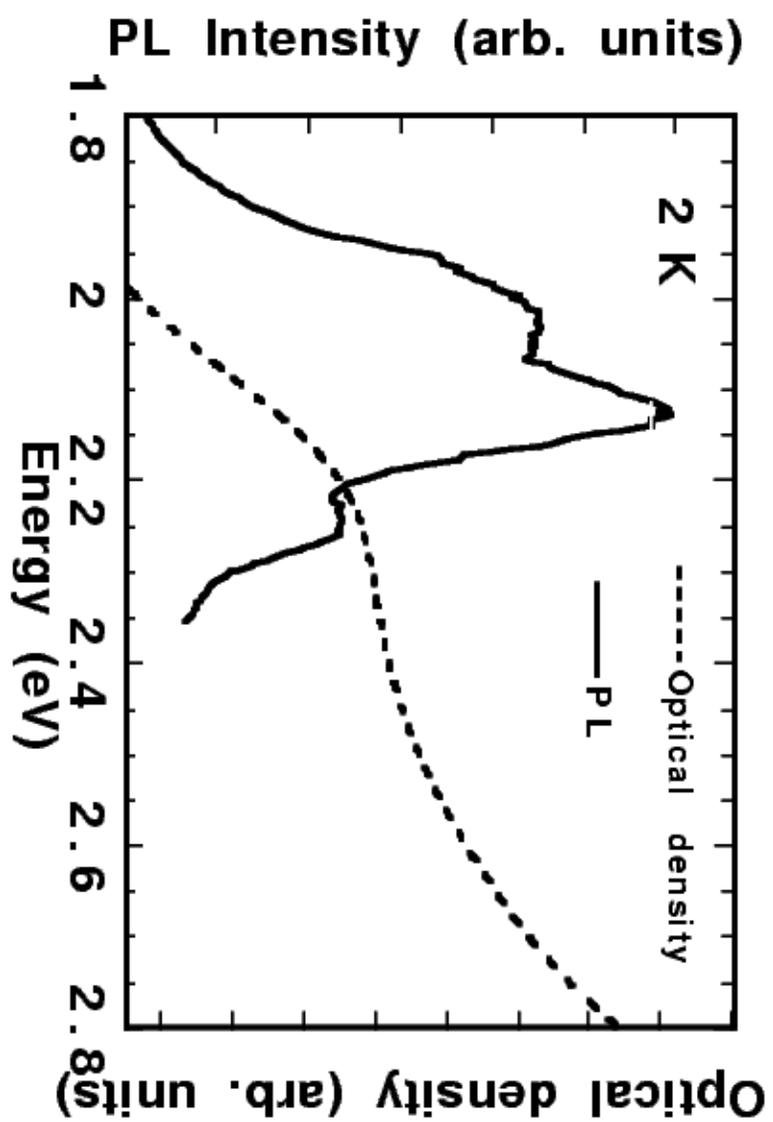
\begin{figure}[b]

\caption{\label{fig1}Low temperature absorption (dashed line) and non-resonant
PL spectra (solid line).}
\end{figure}

Low-temperature absorption and non-resonant PL spectra are represented
in Fig. \ref{fig1}. The absorption spectrum show a very broad maximum
at an energy of \( 2.25 \) eV. Three maxima are clearly observed
in the non-resonant PL spectrum. The maximum of the highest energy
PL band matches very well the first energy maximum of the absorption
spectrum indicating the intrinsic origin of this PL band. The two
lowest energy maxima of PL spectrum should be attributed to the recombination
from states in the energy band gap.

By tuning the laser energy to the red side of the lowest energy absorption
band, one excites only the lowest energy transitions of the biggest
nanocrystals. Fig. \ref{fig2} shows a spectrum obtained in this condition.
The spectrum (resonant PL) reveals fine structures which were not
present in the emission spectrum from the whole distribution (non-resonant
PL). Three peaks are superimposed on the high-energy side of the broad
band peaked at \( 2.125 \) eV (see Fig. \ref{fig1}). The first peak
is shifted with respect to the energy laser excitation (\( 2.236 \)
eV) to the red by about \( 15 \) meV. The other two peaks are the
1LO and 2LO phonon replica of the first peak. When the excitation
energy is changed from \( 2.193 \) to \( 2.306 \) eV the red shift
of the first peak increases from \( 13 \) meV up to \( 20 \) meV
(with an accuracy equal to \( \pm 1 \) meV).

Previous theoretical studies of the exciton structure based on the
effective mass approximation (EMA),\cite{er96} pseudopotential,\cite{ff99}
and tight-binding methods.\cite{lp98, pb01} have been mostly devoted
to CdSe NC's, where the crystal-field term also plays an important
role. Here we apply a recently developed symmetry-based TB theory
\cite{pb99} to zincblende CdTe NC's. It is based on the semi-empirical
\( sp^{3}s^{*} \) TB model of Vogl \emph{et al.} \cite{vh83} extended
to include the spin-orbit interaction, which accounts for the main
features of the bulk semiconductor band structure. We first deduce
the complete single-particle spectrum in a QD of \( T_{d} \) symmetry
by using group-theoretical methods.\cite{pb99} The TB parameters
for CdTe in the present paper are taken from Ref. \onlinecite{be97}:
\( E_{s,a}=-9.86 \), \( E_{p,a}=-0.04 \), \( E_{s,c}=0.4 \), \( E_{p,c}=5.26 \),
\( V_{s,s}=-3.71 \), \( V_{x,x}=1.22 \), \( V_{x,y}=4.12 \), \( V_{s,p}=0.54 \),
\( V_{p,s}=-4.81 \), \( E_{s^{*},a}=7.0 \), \( E_{s^{*},c}=8.50 \),
\( V_{s^{*},p}=2.46 \), \( V_{p,s^{*}}=-0.67 \), \( \lambda _{a}=0.3285 \)
and \( \lambda _{c}=0.0591 \) eV. Where \( E_{ib} \) are the effective
atomic energy levels for cations (\( b=c \)) and anions (\( b=a \)),
\( V_{ij} \) are the hopping parameters between the \( i \) and
\( j \) orbitals on nearest-neighbor atoms and \( \lambda _{b} \)
are the spin-orbit coupling parameters. We consider quasi-spherical
crystallites of zincblende structure ranging from \( 17.81 \) \AA{}
(\( 87 \) atoms) to \( 58.68 \) \AA{} (\( 3109 \) atoms) in diameter.
The dangling bonds are saturated by hydrogen atoms with the H energy
level at \( E_{s,H}=-10 \) eV. The hydrogen-to-atom bond lengths
are chosen so that the surface states have completely disappeared
from the gap and relevant states below and above it. In this work
the cation-hydrogen and anion-hydrogen bond lengths have been fixed
at \( d_{Cd-H}=0.81 \) and \( d_{Te-H}=0.77 \) \AA{} respectively.
A more detailed discussion on the dangling bond saturation in CdTe
nanocrystals was presented previously.\cite{pb99} 

The TB Hamiltonian is written in a block-diagonal form by using a
symmetrized basis corresponding to the double-valued representations
\( \Gamma _{6} \), \( \Gamma _{7} \) and \( \Gamma _{8} \) of the
tetrahedral group \( T_{d} \). An exact diagonalization then yields
the one-particle QD eigenstates \( \phi _{i} \), which can be finally
written as \begin{equation}
\label{state}
\phi _{i}({\textbf {r}})=\sum _{R,k,m}C_{R,k,m}^{i}u_{m}^{k}({\textbf {r}}-{\textbf {R}}),
\end{equation}
 where \( {\textbf {R}} \) denotes the atomic site and \( u^{k}_{m} \)
represent the spin-orbit coupled symmetrized atomic orbitals, with
\( k=6,7,8 \) for \( \Gamma _{6} \), \( \Gamma _{7} \), \( \Gamma _{8} \)
representations respectively.  Let us summarize the essential features
of our results. The two highest occupied valence levels are near each
other in energy and belong to the \( \Gamma _{8} \) (fourfold degenerate)
symmetry. Additionally, for the three biggest NC's studied here, these
two levels are well separate in energy from the others (see Table
\ref{table1}). The dominant orbital symmetry of the highest level
is \( \Gamma _{5} \) for the NC sizes considered here and the nearest
level presents also a non-zero \( \Gamma _{5} \) orbital contribution.
This is in contrast with the EMA results where one of the two highest
valence levels is dipole allowed and the other is forbidden.\cite{rl96}
The symmetry of the lowest conduction state is \( \Gamma _{6} \)
and almost pure \( \Gamma _{1} \) in its orbital part. Furthermore,
the next conduction level is well separated in energy. That means
that the lowest energy part of optical transitions will be driven
essentially by the two \( \Gamma _{8} \) valence quadruplets and
the \( \Gamma _{6} \) conduction doublet (see Table \ref{table1}). 

From the single-particle states we can now write the exciton states
in terms of Slater determinants. Formally, \( |e\rangle =\sum _{v,c}C_{v,c}|v,c\rangle  \),
where \( |v,c\rangle =a_{c}^{\dagger }a_{v}|g\rangle  \) and the
\( a_{c}^{\dagger } \)(\( a_{v} \)) is the creation(annihilation)
operator for a conduction (valence) electron and \( |g\rangle  \)
represents the many-body ground state corresponding to the fully occupied
valence bands. When the \( e-e \) Coulomb interaction is introduced
the matrix elements of the total Hamiltonian can be written as, \begin{equation}
\label{hamiltonian_complete}
H_{vc,v^{\prime }c^{\prime }}=(\varepsilon _{c}-\varepsilon _{v})\delta _{vv^{\prime }}\delta _{cc^{\prime }}-J_{vc,v^{\prime }c^{\prime }}+K_{vc,v^{\prime }c^{\prime }},
\end{equation}
 with \( J \) and \( K \) representing the direct and exchange terms
respectively. When \( J \) and \( K \) are written in terms of the
QD single-particle states in Eq. (\ref{state}), each of them contain
many direct and exchange integrals of atomic orbitals. After standard
simplifying approximations (see Refs. \onlinecite{pb01} and \onlinecite{lp98}
for details), the unscreened on-site Coulomb and exchange integrals
for anion (cation) are assumed to be \( U_{coul}=13(10.5) \) eV and
\( U_{exch}=1(0.5) \) eV respectively. These values follow roughly
those obtained for CdSe. \cite{lp98, pb01} The values for the Te
atom have been calculated by means of a simple scaling. Finally, the
integrals are screened in the Coulomb terms \( J \), but left unscreened
up to the nearest neighbors in the exchange terms \( K \). The on-site
screening factor is taken to be \( 0.4 \) and \( 0.5 \) for cation
and anion respectively. The nearest-neighbor exchange integrals are
assumed one tenth of the on-site ones. The size dependent permittivity
appearing in \( J \) and \( K \) follows the model given by Wang
and Zunger \cite{wz96} for CdSe NC's. In particular, the size dependent
permittivity in the five NC's analyzed here with diameters \( 17.81 \),
\( 24.95 \), \( 32.43 \), \( 43.02 \) and \( 58.68 \) \AA{} is
given by \( 6.14 \), \( 6.64 \), \( 7.09 \), \( 7.68 \) and \( 8.54 \)
respectively.

In order to derive the fine structure of the lowest-energy interband
transitions, we diagonalized the full exciton Hamiltonian in a subspace
of progressively increasing size, with as many valence and conduction
states as necessary to reach numerical convergence. We needed up to
18 valence and 14 conduction states respectively for achieving a convergence
to 0.1 meV for the energy shifts of the eight lowest states. 

The most important contributions to the lowest part of the exciton
spectrum come from the \( 16 \) Slater determinants which correspond
to the direct products of the highest two \( \Gamma _{8} \) valence
states and the the lowest \( \Gamma _{6} \) conduction states. When
the Hamiltonian diagonalization is restricted to the Hilbert space
spanned by these \( 16 \) states the results are close to those reached
at the numerical convergence. The essential physics of the lowest
excitations can be extracted, however, from the eight lowest exciton
states: The symmetry structure of these levels is not changed after
convergence. We find an optically inactive doublet ground state (\( \Gamma _{3} \)
symmetry). The first excited state is a forbidden triplet (\( \Gamma _{4} \)),
which is almost degenerate with the doublet. Finally, the second excited
state is an optically active triplet (\( \Gamma _{5} \)). It is interesting
to note that the above level scheme indeed corresponds to the splitting
expected from the electron-hole exchange interaction. 

The absorption gap is obtained numerically from the first allowed
energy level in the excitonic space, the \( \Gamma _{5} \) triplet.
The numerical results along with the collected experimental data are
shown in Fig. \ref{fig3}.

In our resonant PL experiment the laser excitation creates an optically
active exciton state (the \( \Gamma _{5} \) triplet). After relaxation,
the luminescence is emitted by recombination from the exciton ground
state (the \( \Gamma _{3} \) doublet). The resonant Stokes shift,
\( \Delta _{exch} \), is obtained from the energy difference between
these two levels. The theoretical results along with the experimental
data from the shift between the excitation energy and the luminescence
peak are presented in Fig. \ref{fig4} and show an excellent agreement.
By looking the Table \ref{table1}, it is easy to verify that the
observed Stokes shift of \( 15 \) meV cannot be interpreted in terms
of the energy separation between the highest two valence levels.

When we plot the calculated Stokes shift against the NC diameter the
results are rather close to those from the EMA calculations of Efros
\emph{et al. \cite{er96}} However, \emph{}a more recent EMA theory
\cite{gi98} which includes the long-range part of the electron-hole
exchange interaction, yields larger Stokes shift values. \cite{ll01}

In conclusion, we report the observation of a sizeable resonant Stokes
shift in the photoluminescence from CdTe nanocrystals. The experimental
data show an excellent agreement with the exciton fine structure calculated
in a TB theory which also accounts for the size dependence of the
absorption gap. Moreover, the theoretical analysis indicates that
the observed Stokes shift arises from the electron-hole exchange interaction.

\begin{table}[b]
\begin{tabular}{|c|c|c|c|c|c|}
\hline 
Diameter (\AA)&
\( v_{1} \)&
\( v_{2} \)&
\( v_{3} \)&
\( c_{1} \)&
\( c_{2} \)\\
\hline
\hline 
\( 17.81 \).&
-0.579(8)&
-0.799(8)&
-0.897(8)&
2.747(6)&
2.957(6)\\
\hline 
\( 24.95 \)&
-0.330(8)&
-0.464(8)&
-0.558(8)&
2.466(6)&
2.860(7)\\
\hline 
\( 32.43 \)&
-0.270(8)&
-0.341(8)&
-0.453(8)&
2.097(6)&
2.408(7)\\
\hline 
\( 43.08 \)&
-0.154(8)&
-0.191(8)&
-0.273(8)&
1.972(6)&
2.225(7)\\
\hline 
\( 58.67 \)&
-0.093(8)&
-0.110(8)&
-0.168(8)&
1.823(6)&
1.997(7)\\
\hline
\end{tabular}

\caption{\label{table1}Energy values of the three highest valence and the
two lowest conduction levels in eV. The symmetries are indicated in
parentheses.}
\end{table}

\newpage

{
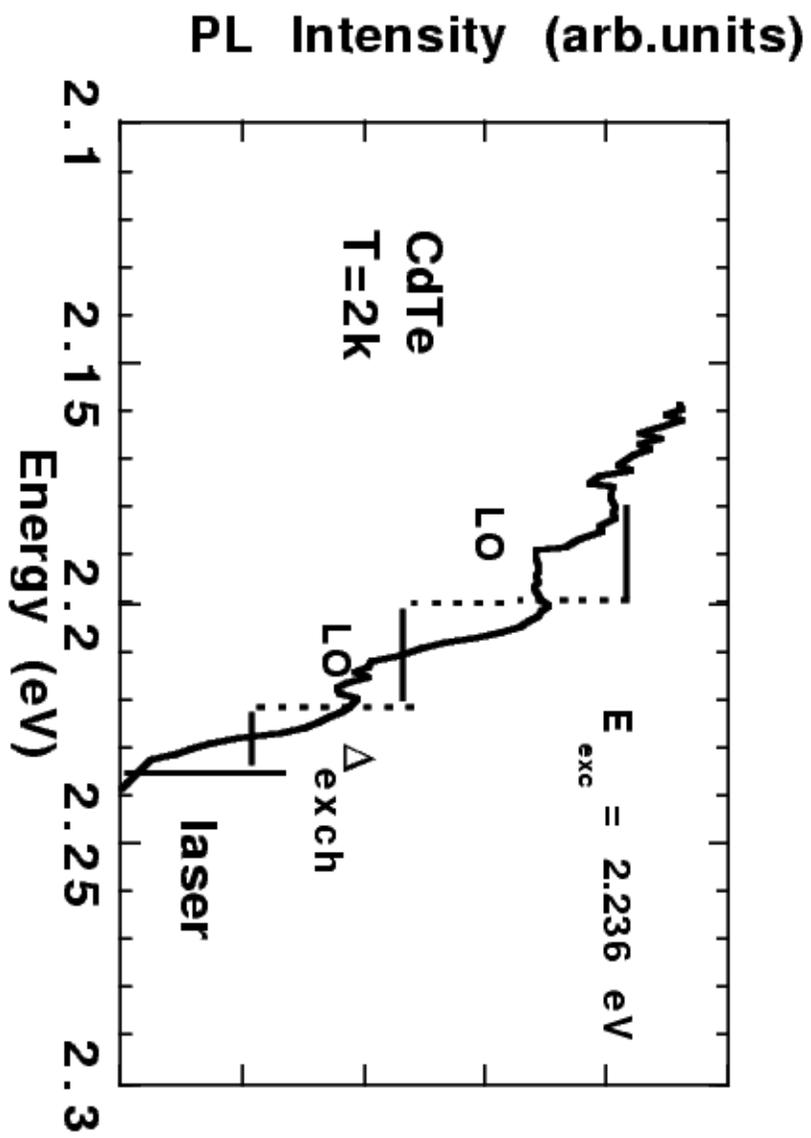
\begin{figure}[b]

\caption{\label{fig2}Photoluminescence spectrum from selective excitation
at \protect\( 2.223\protect \) eV.}
\end{figure}
\par}

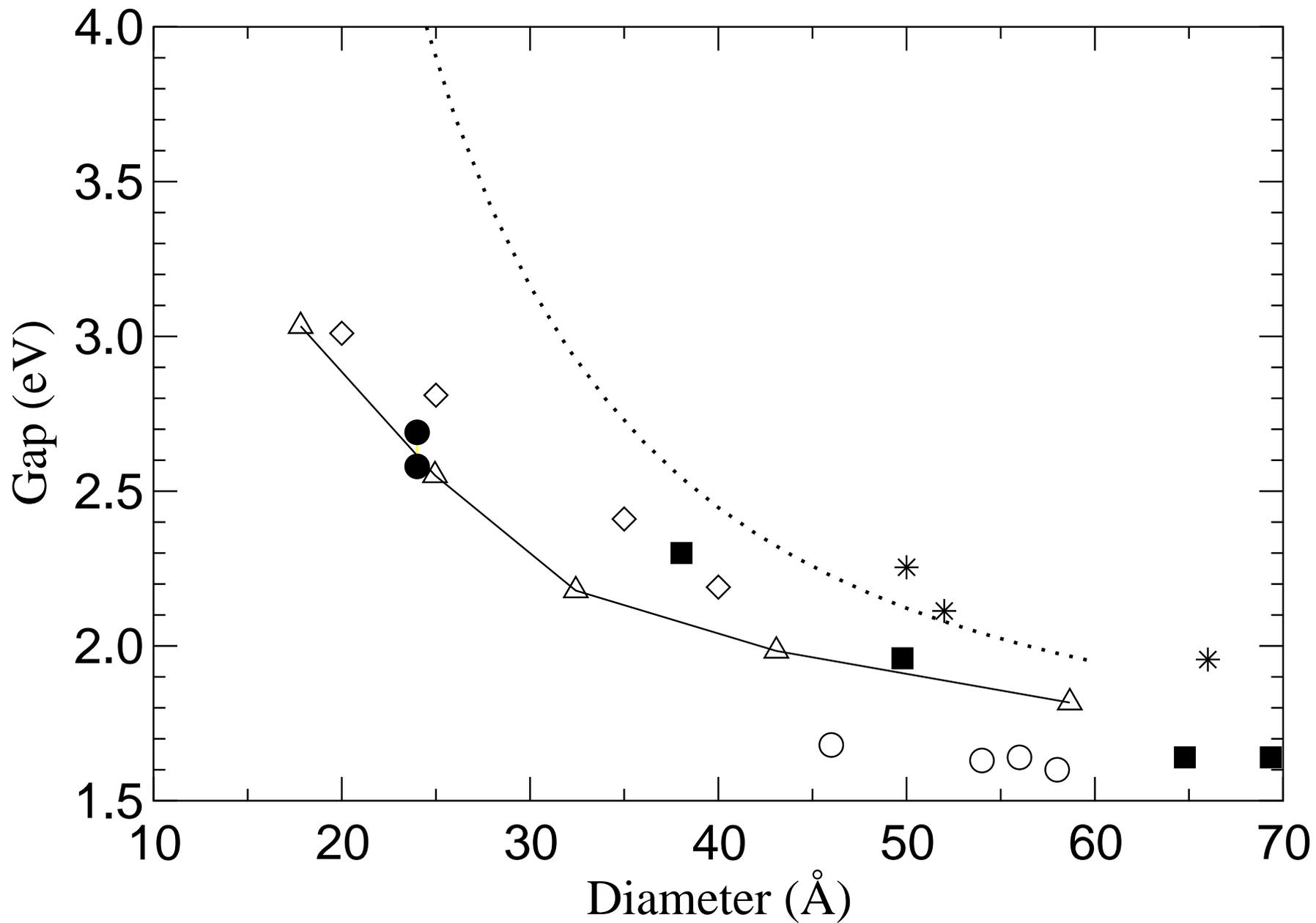
\begin{figure}[b]

\caption{\label{fig3}Absorption band gap versus QD diameter. The experimental
data were taken at different temperatures. The data from Ref. \protect\onlinecite{rm93}
(open diamonds) and Ref. \protect\onlinecite{mh97} (closed squares)
were measured at the room temperature. The data from Ref \protect\onlinecite{ms97}
(stars) were measured in superfluid helium at \protect\( 2\protect \)
K. The values from Ref. \emph{}\protect\onlinecite{ar00} were taken
at temperatures higher than room temperature (open circles). Finally
the sample studied in Ref. \protect\onlinecite{ks99} (closed circles)
was measured at 77 K (higher value) and room temperature (lower value).
Our theoretical TB values scale (open triangles) as a function of
the diameter as \protect\( 1/D^{1.42}\protect \), while the EMA\cite{rl96}
gap (dotted line) varies as \protect\( 1/D^{2}\protect \).}
\end{figure}

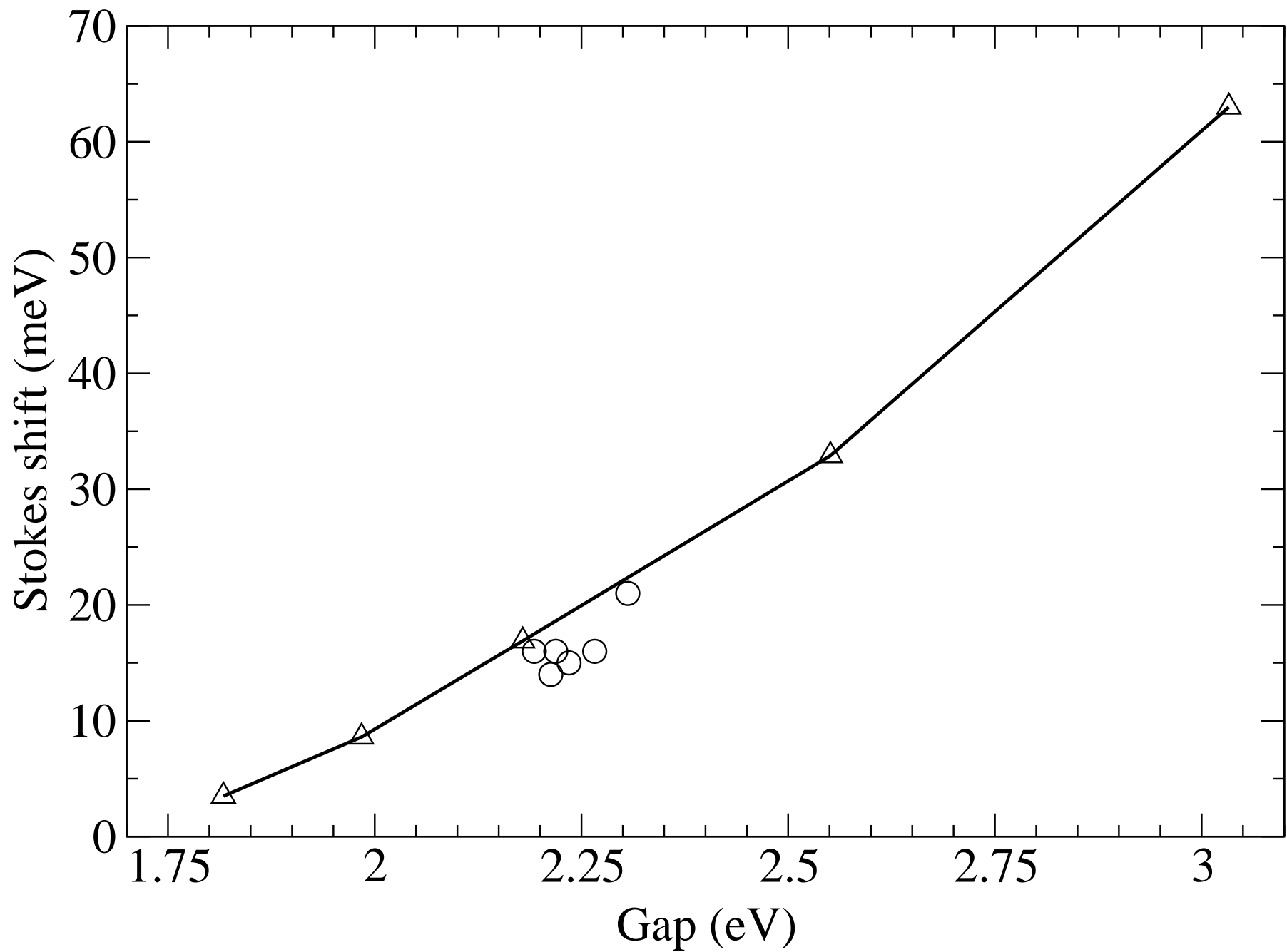
\begin{figure}[b]

\caption{\label{fig4} The resonant Stokes shift is plotted against the absorption
gap. The experimental values (open circles) and the calculated ones
(open triangles) are compared. }
\end{figure}

\end{document}